\documentclass[structabstract]{aa}
\usepackage{amssymb}
\usepackage[usenames, dvipsnames]{color}
\usepackage{graphicx}
\usepackage{natbib}
\usepackage[rightcaption]{sidecap}
\usepackage{txfonts}
\usepackage{units}
\usepackage{url}
\usepackage{pifont}
\usepackage[modulo, switch]{lineno}
\graphicspath{ {images/} }
\usepackage{ulem}

\definecolor{blue}{rgb}{0.00, 0.00, 1.00}
\definecolor{red}{rgb}{0.86, 0.08, 0.24}
\definecolor{orange}{rgb}{1.00, 0.55, 0.00}
\definecolor{darkblue}{rgb}{0.00, 0.00, 0.55}
\definecolor{green}{rgb}{0.00, 0.39, 0.00}
\definecolor{pink}{rgb}{1.000000,0.078431,0.576471}

\begin{document} 

%###############################################################################
%#
%#    TITLE
%#
%###############################################################################

\title{Properties of the inner penumbral boundary and 
       temporal evolution of a decaying sunspot}
\titlerunning{Properties of the inner penumbral boundary and 
       temporal evolution of a decaying sunspot.}

\author{%
    M.\ Benko\inst{1},
    S. J.\ Gonz{\'a}lez Manrique\inst{1},    
    H.\ Balthasar\inst{2}, 
    P.\ G{\"o}m{\"o}ry\inst{1},
    C.\ Kuckein\inst{2} and
    J.\ Jur\v{c}\'{a}k\inst{3}
    }

\authorrunning{Benko Martin et al.}
   
\institute{%
    $^1$ Astronomical Institute, Slovak Academy of Sciences, 
         05960 Tatransk\'{a} Lomnica, Slovak Republic\\
    $^2$ Leibniz-Institut f{\"u}r Astrophysik Potsdam (AIP),
         An der Sternwarte 16, 
         14482 Potsdam, Germany\\
    $^3$ Astronomical Institute of the Academy of Sciences, 
         Fri\v{c}ova 298, 25165 Ond\v{r}ejov, Czech Republic\\     
    \email{mbenko@ta3.sk}}

\date{Received September 21 2018; accepted October 29 2018}
 
\abstract
% Context
{It was empirically determined that the umbra-penumbra boundaries of stable sunspots are characterized by a constant value of the vertical magnetic field. }
% Aims
{We analyzed the evolution of the photospheric magnetic field properties of a decaying sunspot belonging to NOAA 11277 between August 28 - September 3, 2011. The observations were acquired with the spectropolarimeter on-board of the Hinode satellite. We aim to proof the validity of the constant vertical magnetic-field boundary between the umbra and penumbra in decaying sunspots. }
%Methods
{A spectral-line inversion technique was used to infer the magnetic field vector from the full-Stokes profiles. 
In total, eight maps were inverted and the variation of the magnetic properties in time were quantified using linear or quadratic fits. }
% Results
{We found a linear decay of the umbral vertical magnetic field, magnetic flux, and area. The penumbra showed a linear increase of the vertical magnetic field and a sharp decay of the magnetic flux. In addition, the penumbral area quadratically decayed. 
The vertical component of the magnetic field is weaker on the umbra-penumbra boundary of the studied decaying sunspot compared to stable sunspots. Its value seem to be steadily decreasing during the decay phase. Moreover, at any time of the shown sunspot decay, the inner penumbra boundary does not match with a constant value of the vertical magnetic field, contrary to what was seen in stable sunspots.}
% Conclusions
{During the decaying phase of the studied sunspot, the umbra does not have a sufficiently strong vertical component of the magnetic field and is thus unstable and prone to be disintegrated by convection or magnetic diffusion. No constant value of the vertical magnetic field was found for the inner penumbral boundary.}


\authorrunning{Benko Martin et al.}

\maketitle
   
%###############################################################################
%#
%#    INTRODUCTION
%#
%###############################################################################

\section{Introduction}\label{SEC1}
\par

Active  regions  (AR)  are  manifestations  of large-scale 
magnetic fields in the solar atmosphere. The largest magnetic structures in ARs are sunspots. 
They are the longest-known manifestation of solar activity. Moreover, sunspots are usually long-lasting, 
existing from a few days up to several months \citep{1951ASPL....6..146P}. 
Sunspots are dark features with a strong magnetic field \citep{1908ApJ....28..315H}. 
A large number of analyses described the global properties of the magnetic field in sunspots, for an overview see the reviews by \cite{Solanki2003} and  
\cite{Borreroichimoto2011}.

Each sunspot is characterized by a dark core, the umbra, and a filamentary penumbra that is 
surrounding the dark core. The presence of a penumbra distinguishes sunspots from pores. 
The magnetic field of the umbra is stronger and more vertical 
than in the penumbra.
There is a sharp intensity boundary between the umbra and penumbra of a sunspot. As shown by \citet{Jurcak_etal2018},
the intensity threshold of 50\% of the quiet-Sun intensity in the visible continuum can be used to define the umbral
boundary.

To understand the physics of sunspots, one has to study their temporal evolution. 
Formation and decaying phases play an important role in sunspot evolution. 
We refer to \citet{vmp2002} for a review of decaying sunspot evolution. 
For a long time, it was only possible to study 
the decay of the morphological changes of the area of sunspots because of the lack of inversion codes to 
interpret full-Stokes measurements, although the magnetic flux is the more 
important parameter. \citet{bumba1963} found a linear decrease of the 
area of the sunspot with time. 
\citet{vmp_etal1993} confirmed this linear decay, but obtained a 
different coefficient. In contrast, \citet{petrovay1997} found a parabolic 
decay with a rate proportional to $\sqrt{A(t)}$, where $A$ is the area of the spot. 
Linear decay rates of the areas of 32 sunspots were found by \citet{Chapman2003}. 
\citet{BandS2005} investigated the Greenwich sunspot group record, but they were not able to distinguish between
a linear and a quadratic decay law. \citet{HandC2008} published an almost constant decay rate of 
3.65$\,\mu$Hemispheres\,day$^{-1}$. \citet{Gafeira2014} studied four sunspots and found an approximately linear decay 
of the areas similar for umbra and penumbra. In both studies, the decay rates are different for individual spots.
Case studies resulting in a linear decrease of the magnetic flux during the decay phase were presented by 
\citet{Dengetal2007} and \citet{Vermaetal2016}. In the latter publication, the development of the area is 
non-monotonic.
\citet{Sheeley2017} investigated the development of the magnetic flux in 36 sunspots, but not all of 
them were in the decaying phase. Some spots showed a nearly linear decay, but they found also indications of 
a bursty decay.
In a 100 hour numerical simulation of a mature sunspot, \citet{Rempel2015} found a linear decay 
of the magnetic flux in the umbra for the last 80 hours. The penumbral magnetic flux remained almost 
constant during this period.

\citet{Jurcak2011} studied the changes of the magnetic field strength
and inclination at the boundaries between the umbra and penumbra (UP). The author
found that the vertical component of the magnetic field $B_z$ was constant, although  
slight changes along the boundary can happen. At the same time, the magnetic field 
strength and inclination vary significantly along the UP boundary. 
\citet{Jurcak_etal2015} found that the umbral areas that have vertical magnetic fields
$B_z$ lower than the abovementioned constant value are colonized by the
penumbra in a forming sunspot. This scenario is a possible mechanism 
to generate orphan penumbrae \citep{Jurcak_etal2017} when the whole umbra 
has $B_z$ lower than the critical value. \citet{Jurcak_etal2018} carried 
out a statistical analysis of the magnetic field  properties of more than 
100 stable umbral cores. The authors narrowed down the critical value of the vertical 
component of the magnetic field on the 50\% intensity boundary to be 1867~G and 
confirmed its invariance. We hereafter call it the \textit{Jur\v{c}\'ak criterion} 
following the definition introduced by \citet{Schmassmann:2018}.

In this paper we study the evolution of a sunspot during its decay phase.
The observations are described in Sect.~\ref{SEC2}. The analysis of the data is
presented in Sect.~\ref{SEC3}. The temporal evolution 
of the physical parameters of the sunspot is discussed in Sect.~\ref{SEC:res_x(t)}. 
The properties of the UP boundaries in the decaying phase of the sunspot is studied
and compared to the constant $B_z$ in stable sunspots (Sect.~\ref{SEC4_4.2}). 
Finally, a discussion and conclusion section are presented in Sect.~\ref{SEC5}.

%###############################################################################
%#
%#    Observations and data reduction
%#
%##############################################################################
\section{Observations and data reduction}
\label{SEC2}

%---- Figure 1 -----------------------------------------------------------------
\begin{figure} [!t]
\centering
\includegraphics[width=\columnwidth]{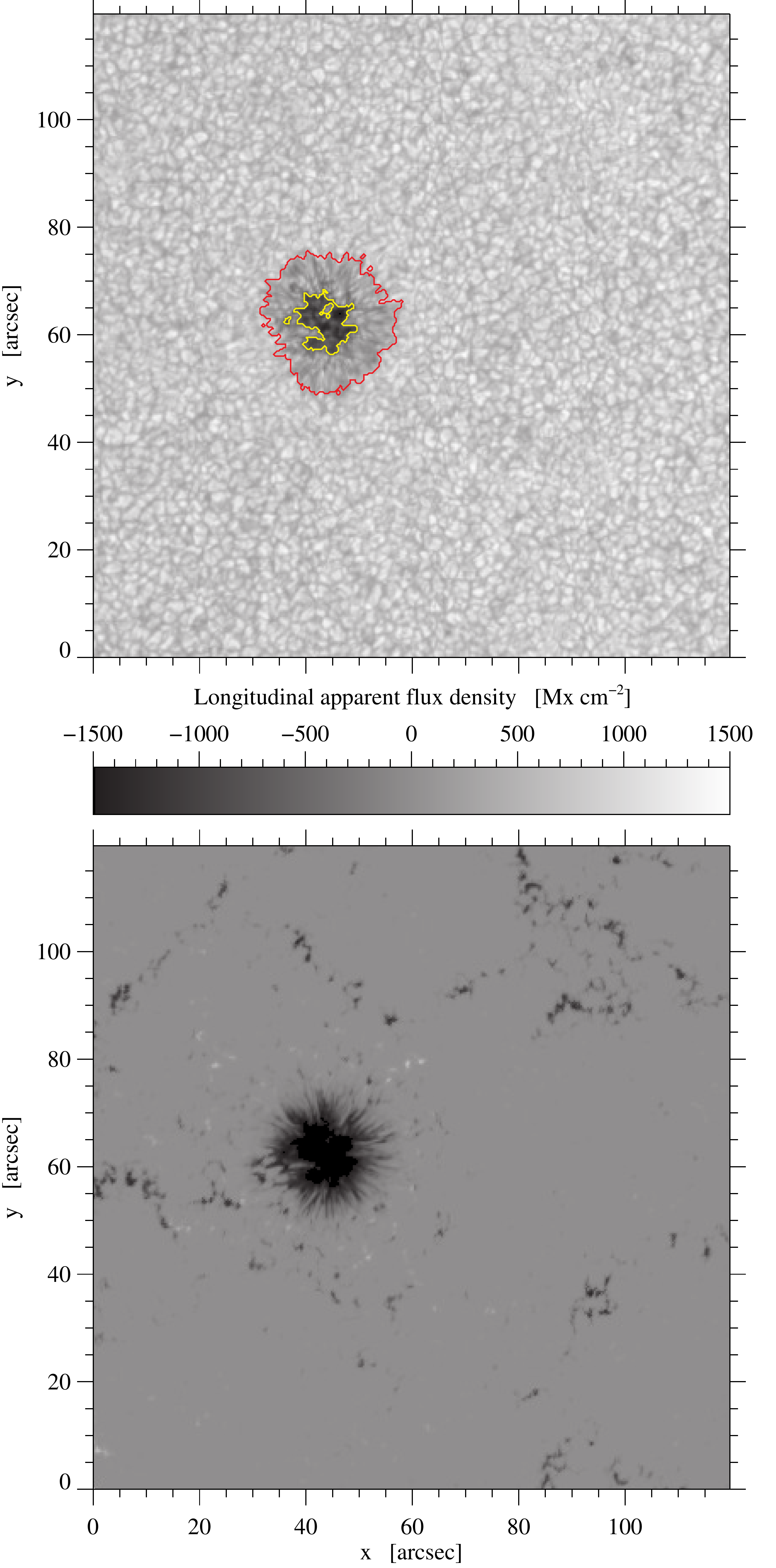}
\caption{Overview of the acquired data. SOT continuum intensity 
          at 07:47:07~UT on 2011 August 31 of the AR
          NOAA 11277 (\textit{top}). Longitudinal apparent flux density
          at the same time (\textit{bottom}). The yellow contour depicts the 
          intensity at $0.5 I_0$ (same as in Figs.~\ref{FIG:jurcak1} and \ref{FIG:jurcak2}). 
          The red contour represents the outer penumbra 
          boundary. It was determined by the horizontal magnetic field  at 
          B$_{hor} \sim 490 G.$
         }
\label{FIG1}
\end{figure}
%------------------------------------------------------------------------------

%--------------Observations-------------------------
A decaying sunspot with negative polarity in AR NOAA 11277, 
was observed on the solar disk between August 28 and September 3 on 2011. 
The observations were acquired with the spectropolarimeter (SP), which is part of the
0.5-meter Solar Optical Telescope \citep[SOT,][]{Ichimoto2008, Tsuneta2008, Suematsu2008},
on-board of the Japanese Hinode satellite \citep{Kosugi2007}.
The data were observed twice a day during the seven days. The recording time 
was approximately between 7:30~UT and 8:00~UT for the first observation of the day
and between 10:00~UT 10:30~UT for second observation. 
We analyze eight SP scans taken during these seven days. 
The active region was located at heliocentric coordinates from $\mu \equiv \cos\theta = 0.87$ 
(first observing day) to $\mu \equiv \cos\theta = 0.75$ (last observing day).
An example of the acquired data is represented in Fig.~\ref{FIG1}.

The SP instrument consists of a Littrow spectrograph,
which measures the full-Stokes parameters of the neutral 
iron lines at \ion{Fe}{i} 6302.5~\AA\ and \ion{Fe}{i} 
6301.5~\AA. The two observed spectral lines are sensitive
to the magnetic field. The Land{\'e} factors 
of these spectral lines are $g=2.5$ and $g=1.7$, respectively. They
are formed in the higher layers of the photosphere. The
\ion{Fe}{i} 6302.5~\AA\ line is formed about 60\,km lower in
the Sun's atmosphere than the \ion{Fe}{i} 6301.5~\AA\ line
\citep{Faurobert09}. For the present analysis, only the photospheric
\ion{Fe}{i} 6302.5~\AA\ line is used because it is more sensitive
to the magnetic field. The purpose of inverting just one line is to save computing time. 
We tested that the results are nearly identical if both lines are inverted.

The region of interest covers a maximum area of $123\arcsec \times 123\arcsec$ 
(the width of the scan varied from day to day with a maximum difference of 
around 8\arcsec, the height of the scan is determined by the SP slit), with 
a spatial scan consisting of 400 steps with a spacing of $0\farcs3$. 
The spatial sampling on the solar surface is of 220~km\,pixel$^{-1}$. In this 
study we only concentrate in a smaller FOV centered on the sunspot.

%---------------data reduction---------------------

The data were downloaded from the Community Spectro-polarimetric
Analysis Center (CSAC) \footnote{http://mlso.hao.ucar.edu/CSAC/sp\_data.php}
in a pre-processed "level 1" format. The data were already calibrated and ready
for scientific purposes. 

Only one map per day is analyzed
because the changes in the sunspot within a few hours are small. We make an exception for 
2011 September 2, where the morphological changes between both maps become evident within 2.5 hours. We use also the second map of this day.

%###############################################################################
%#
%#    Data analysis
%#
%##############################################################################

\section{Data analysis}
\label{SEC3}

%---- Figure 2 -----------------------------------------------------------------
\begin{figure} [!t]
\centering
\includegraphics[width=\columnwidth]{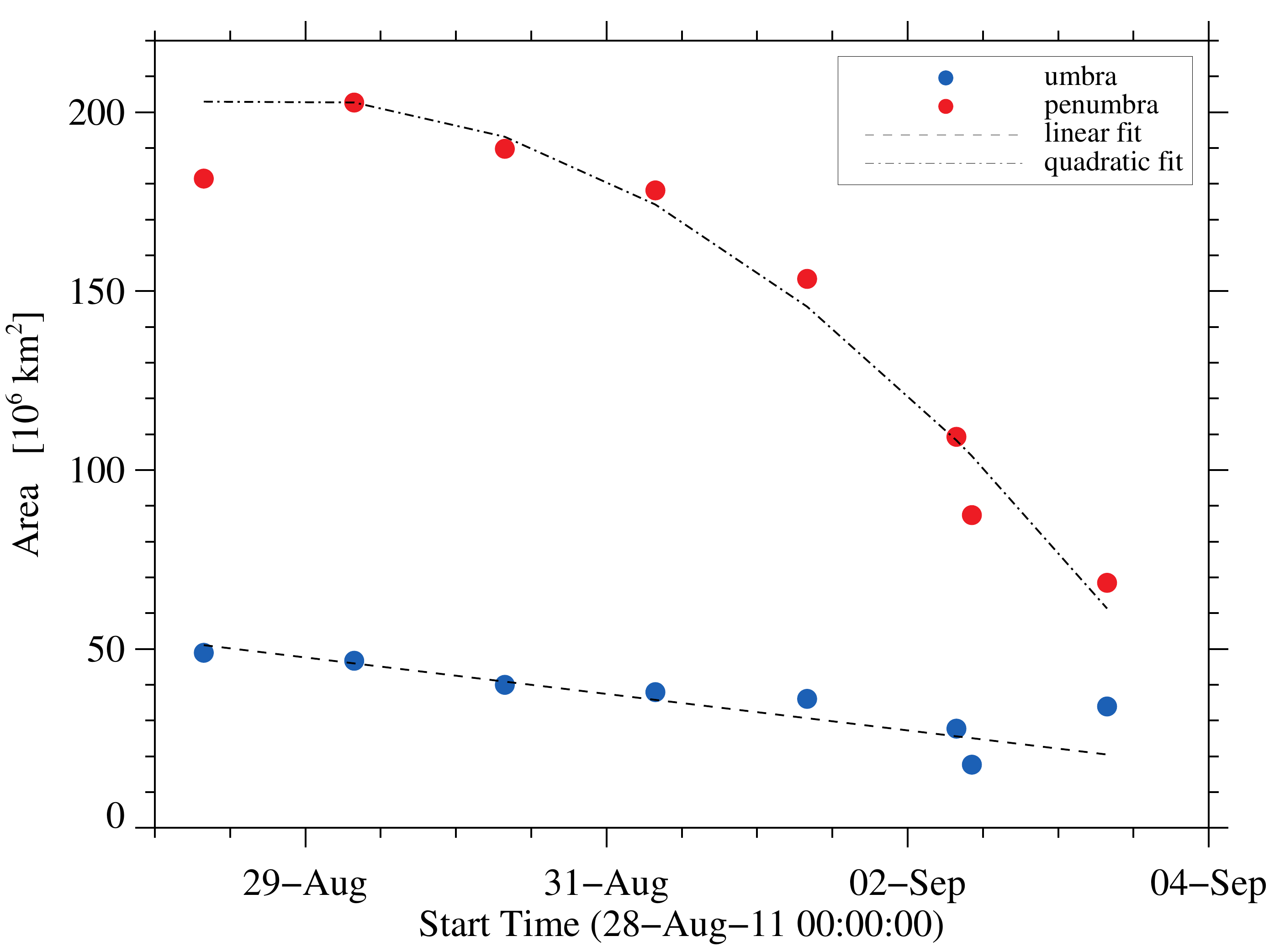}
\caption{Temporal evolution of the area of the umbra (blue dots) and penumbra (red dots) of 
         the sunspot during the observing time with SOT. The dashed line represents the linear 
         fit for the values of the umbra and the dashed-dotted line depicts the linear quadratic fit 
         for the values of the penumbra.
         }
\label{FIG:area(t)}
\end{figure}
%------------------------------------------------------------------------------
%---- Figure 3 -----------------------------------------------------------------
\begin{figure} [t]
\centering
\includegraphics[width=\columnwidth]{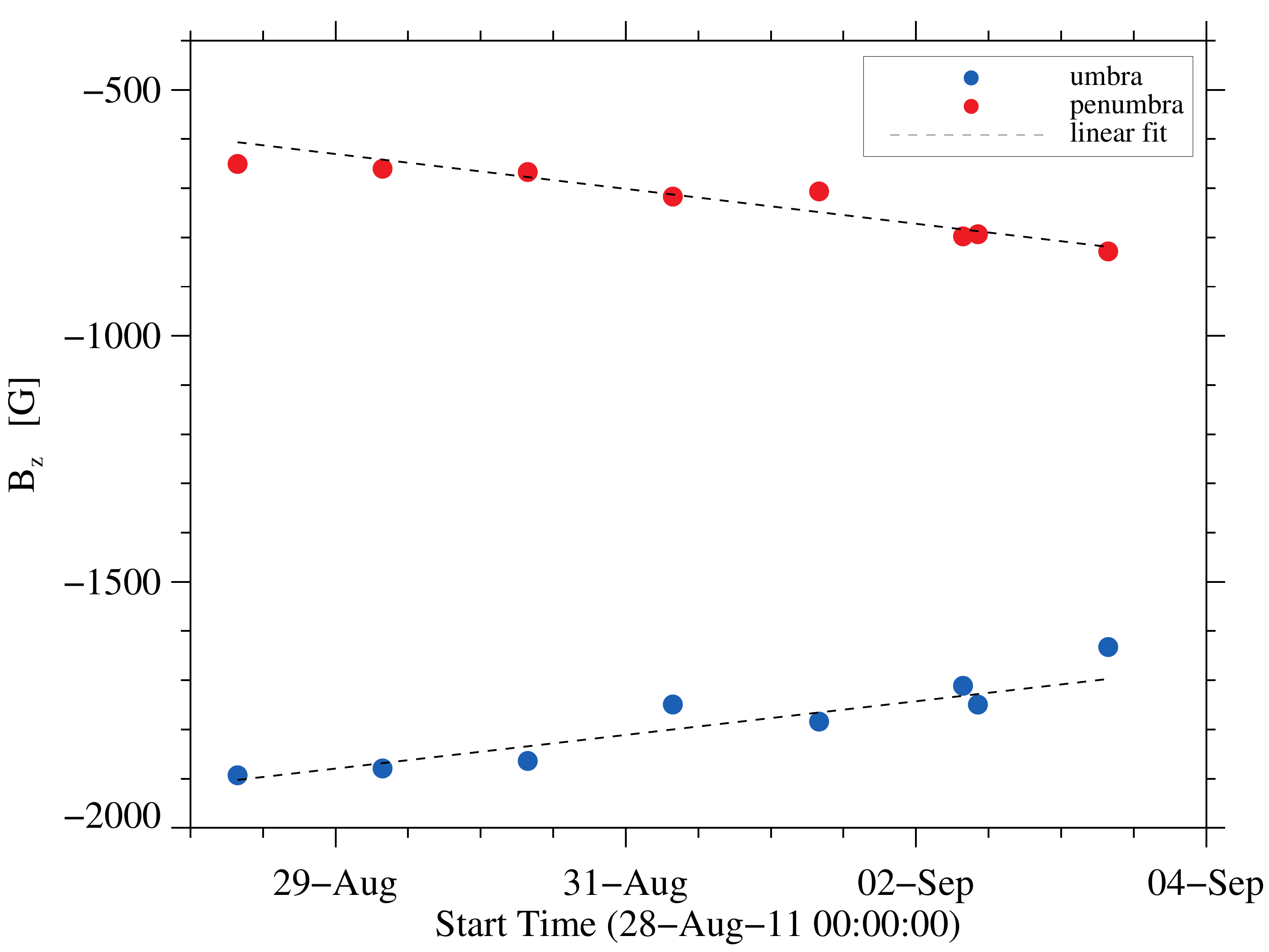}
\caption{Temporal evolution of the mean vertical component of the magnetic field
         $B_z$ calculated in the umbra (blue dots) and the penumbra (red dots). 
         The dashed line represents the linear fit to the dots. }
\label{FIG:Bz}
\end{figure}
%------------------------------------------------------------------------------

The Stokes parameters are inverted using the Stokes Inversion based 
on Response functions \citep[SIR,][]{Ruizcobo2012}.
The code delivers physical parameters such as the magnetic field vector,
the temperature, and the Doppler velocity among others. The analyzed spectral line
is formed in a narrow height range above the surface of the Sun. Therefore, the potential
height gradients of the individual parameters obtained from the inversion are not relevant 
for the present investigation, and we do not take them into consideration for the further 
analysis. Thus, we restrict the inversions to one node for the 
Doppler velocity, magnetic field strength, inclination and 
azimuth, keeping these values height-independent. Only the temperature is 
height-dependent with three nodes. We assume a single atmospheric component.

%---- Figure 4 -----------------------------------------------------------------
\begin{figure} [!t]
\centering
\includegraphics[width=\columnwidth]{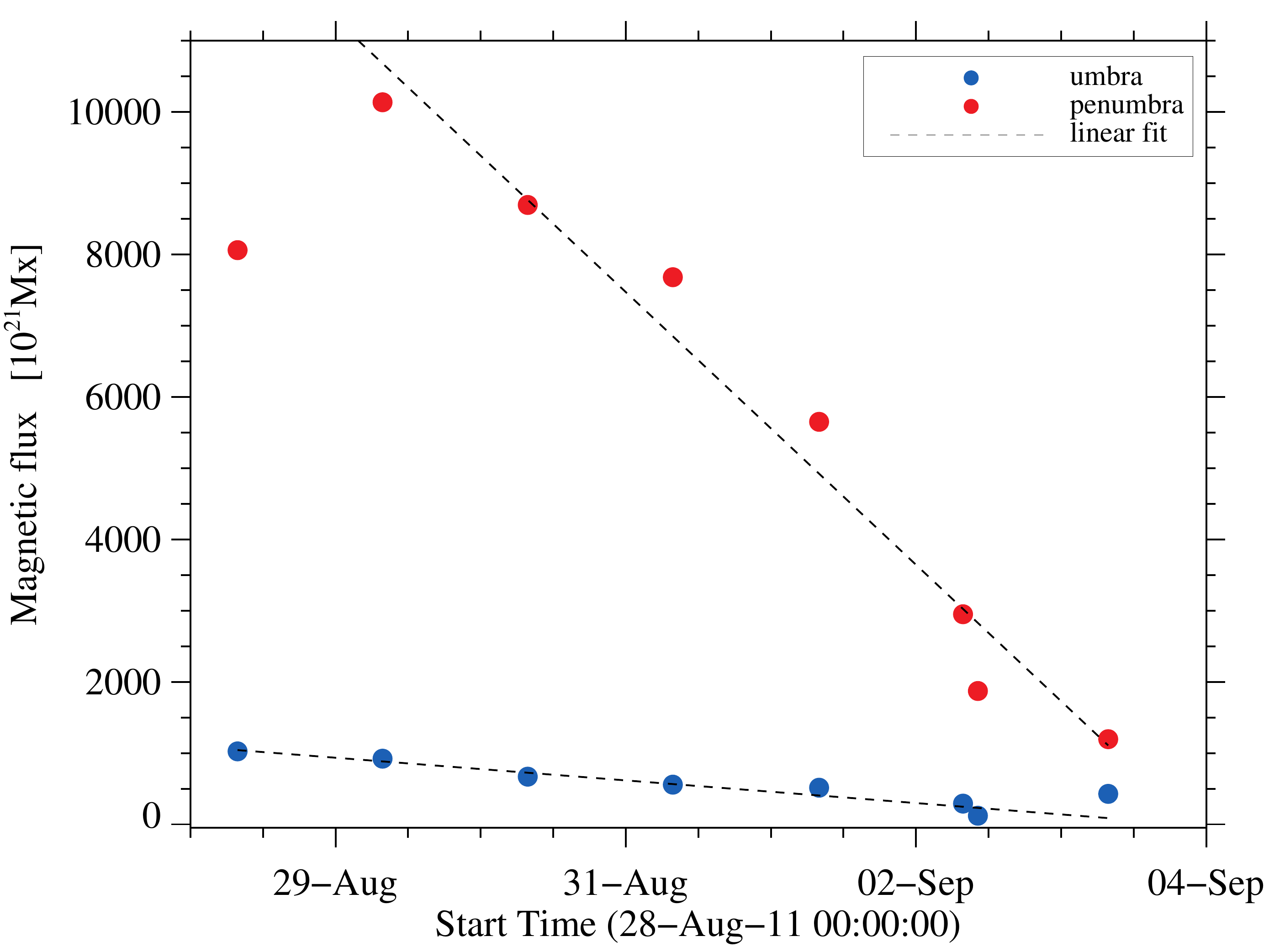}
\caption{Same as Fig. \ref{FIG:Bz} but for the temporal evolution of the total magnetic flux.}
\label{FIG:flux}
\end{figure}
%------------------------------------------------------------------------------
%---- Figure 5 -----------------------------------------------------------------
\begin{figure} [!t]
\centering
\includegraphics[width=\columnwidth]{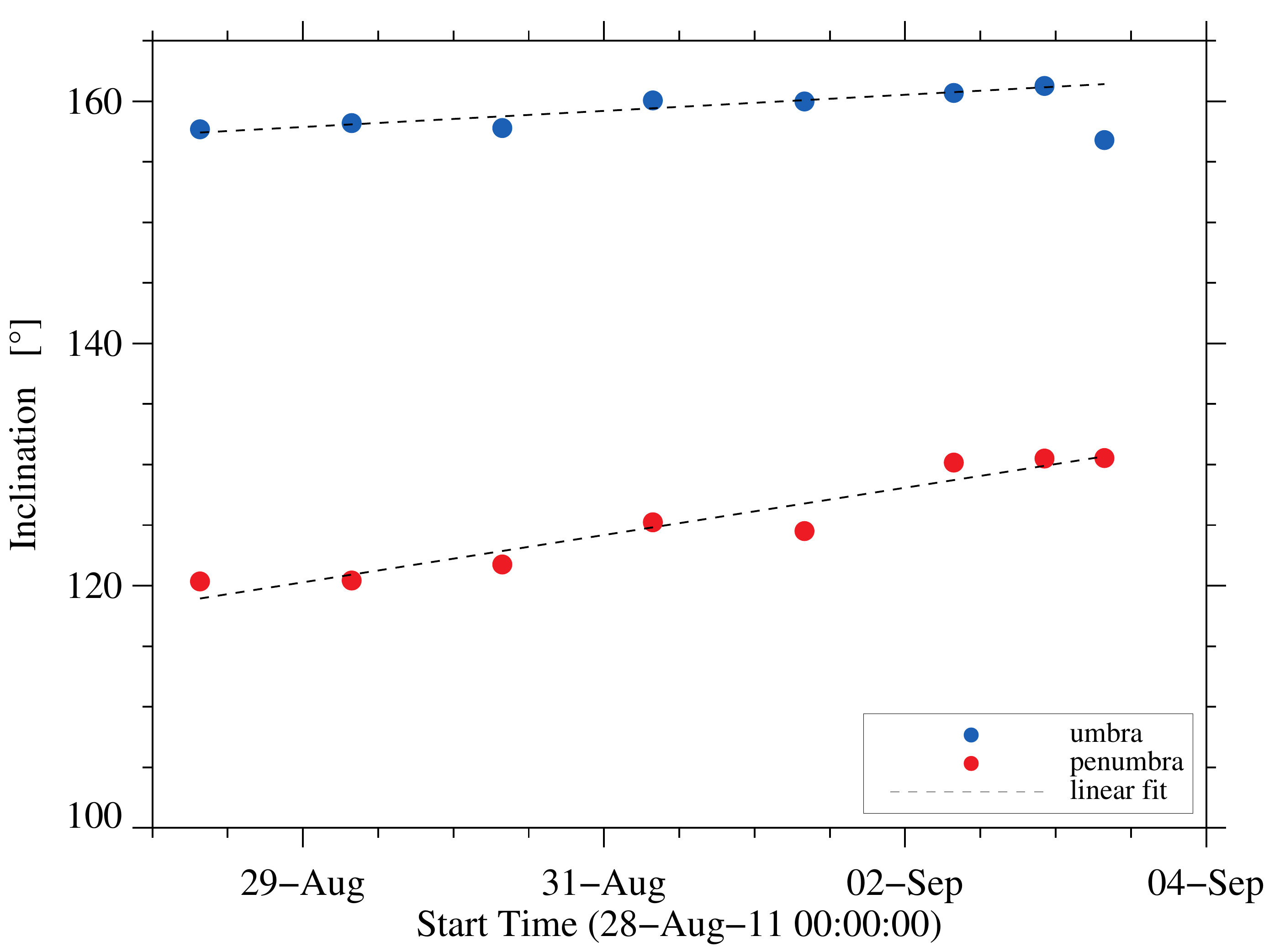}
\caption{Same as Fig. \ref{FIG:Bz} but for the temporal evolution of the inclination.}
\label{FIG:inclination}
\end{figure}
%------------------------------------------------------------------------------

We assume that there is a single azimuth center in the sunspot to solve 
the 180-degree ambiguity \citep[see,][]{BalthasarCEAB2013}.
Such an azimuth center has a radial magnetic field configuration,
and the observed azimuth values for each pixel differ by less 
than $\pm$ 90$^\circ$ from the radial configuration. The last step
is to transform the coordinate system of the magnetic field to the local
reference frame and to correct the geometrical foreshortening of 
the observations taken at different position angles applying a method described by \citep{Verma2012}.

%###############################################################################
%#
%#    Results
%#
%##############################################################################

\section{Results}\label{SEC:results}

%=======================================================================================
%    Temporal evolution of area, vertical magnetic field, magnetic flux, and inclination
%=======================================================================================

\subsection{Temporal evolution of area, vertical magnetic field, magnetic flux, and 
    inclination}
\label{SEC:res_x(t)}

%---- Figure 6 -----------------------------------------------------------------
\begin{figure*} [!t]
\centering
\includegraphics[width=\textwidth]{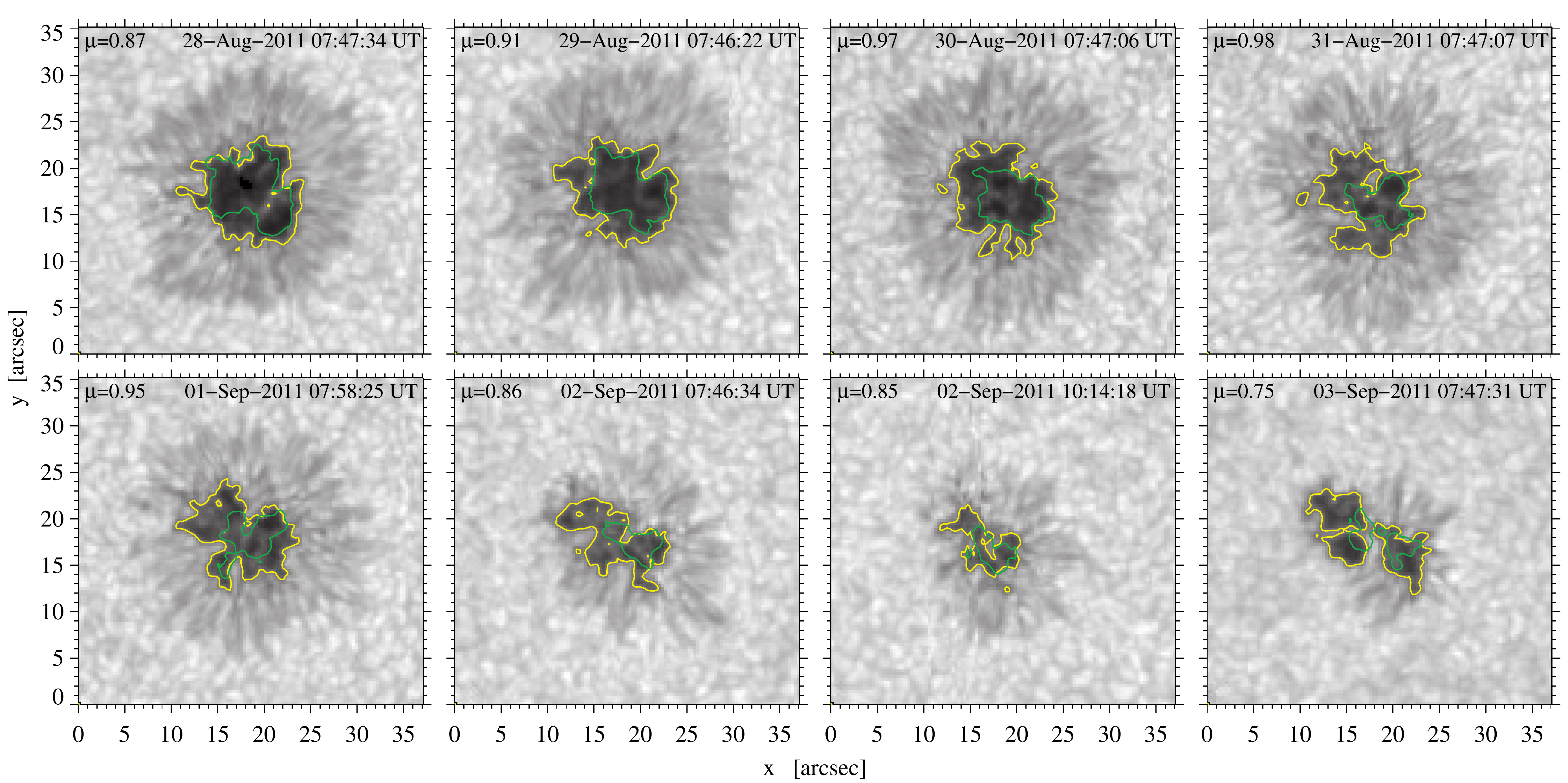}
\caption{Temporal evolution of the sunspot in active region NOAA 11277 during seven 
         days starting at 07:47 UT on 2011 August 28 (\textit{top-left to bottom-right}). All 
         images are normalized to the intensity of the local continuum. The yellow contours
         represent the intensity at $0.5 I_0$. The green contours depict
         the magnetic field component perpendicular to the solar surface ($B_z$) at 1867~G (\textit{Jur\v{c}\'{a}k criterion}). 
         }
\label{FIG:jurcak1}
\end{figure*}
%------------------------------------------------------------------------------

%---- Figure 7 -----------------------------------------------------------------
\begin{figure*} [!t]
\centering
\includegraphics[width=\textwidth]{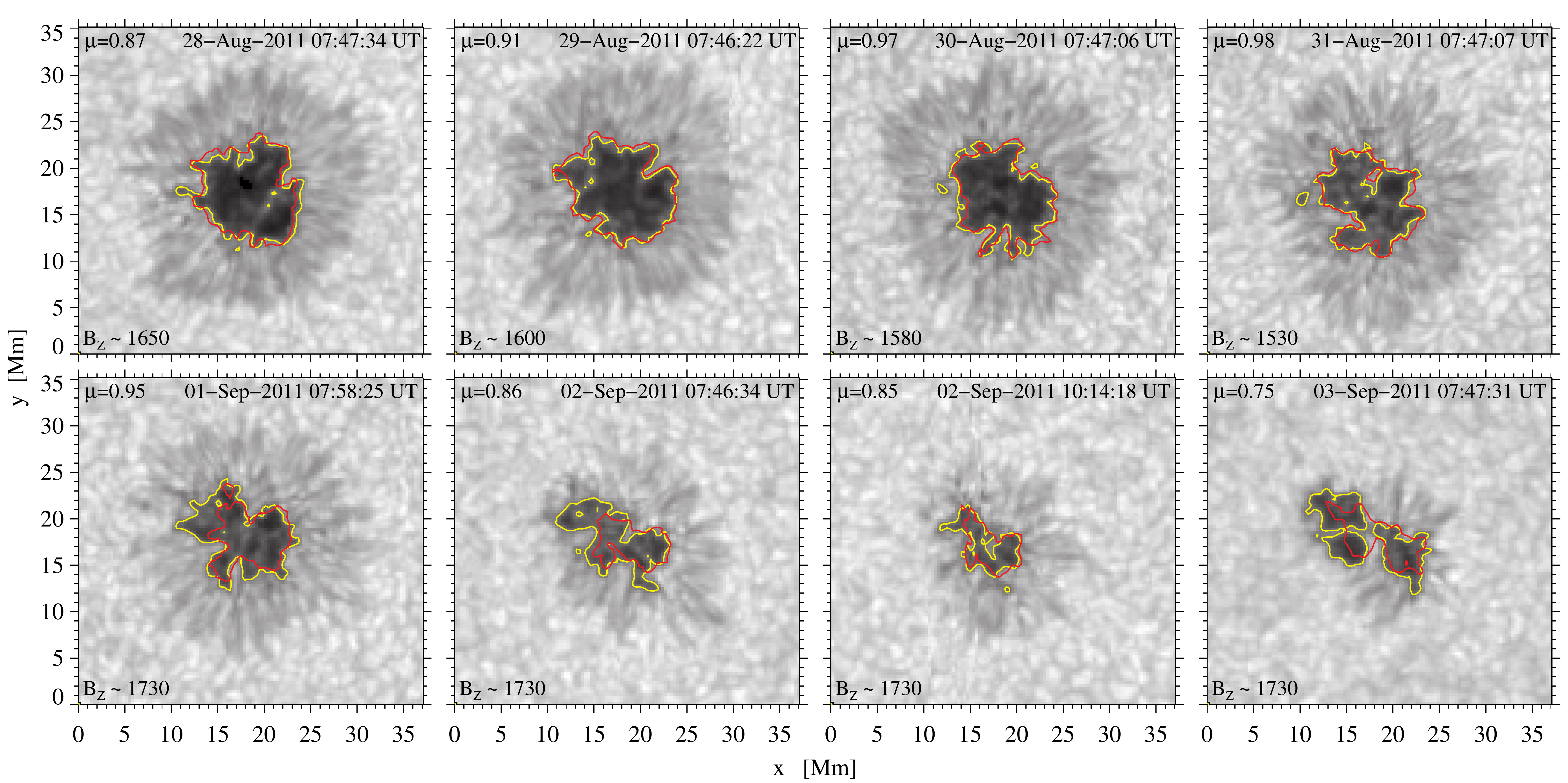}
\caption{Temporal evolution of the same observing days as in Fig.~\ref{FIG:jurcak1} but adding the red countours,            
         which represent the magnetic field component perpendicular to the solar surface $B_z$ that best matches 
         the umbra-penumbra boundary (yellow contours).
 }
\label{FIG:jurcak2}
\end{figure*}
%------------------------------------------------------------------------------

The temporal evolution of the areas of umbra and penumbra is displayed in Fig.~\ref{FIG:area(t)}.
The area of the umbra is determined using the UP boundaries at intensity level $0.5 I_0$, 
where $I_0$ is the average quiet-Sun intensity of the local continuum. The outer penumbra
boundary, necessary to identify the penumbral area, is determined by the horizontal magnetic
field at B$_{hor} = \sqrt{B_x^2 + B_y^2} \sim 490 G$ (as an example see the contours in Fig.~\ref{FIG1}). 
With this magnetic criterion, we obtain a better contour of the outer penumbral boundary than with any intensity
criterion. The magnetic pressure of such a horizontal field would counterbalance the dynamic pressure of rising
convective elements between 2 and 4 km\,s$^{-1}$, depending on the density of matter in the atmosphere outside the
magnetic field and the Wilson depression of the outer penumbra, see also \cite{Wiehr96}.

We apply linear and quadratic fits to the data, but we show the quadratic fit only when there is 
a clear curvature. On September 3, obviously new umbral flux raised (see also Fig.~\ref{FIG:flux}),
therefore we omit this day from the fits for the umbra. On August 28, the penumbral area was 
smaller than on August 29, probably the penumbra was still in the growing phase on this day,
and thus we omit this day from the fits which should represent the decay phase. The same days for
umbra and penumbra are excluded from the fits in Figs.~\ref{FIG:Bz}, \ref{FIG:flux}, and~\ref{FIG:inclination}.

The umbral area decreases linearly during the period of observation. 
For the penumbra we see an accelerated decay, and a parabola fit yields a better representation. 
The penumbra has an area of $200\,{\rm Mm}^2$ at the beginning of the observations. Then the area 
decreases to a value of $70\,{\rm Mm}^2$ on September 3. 
The umbra has a value of $49\,{\rm Mm}^2$ at the beginning of the observations and $18\,{\rm Mm}^2$ at the end of observation.

Fig.~\ref{FIG:Bz} shows the temporal evolution of the vertical component of the magnetic 
field $B_z$ for umbra and penumbra. The absolute umbral value decreases from 1900\,G on August 28 to 
1600\,G on September 02 (keep in mind that the polarity of the spot is negative). As for the area, 
there is no significant deviation from a linear decay. The penumbra has a value of 650\,G at the beginning. 
This value increases with time, and $B_z$ is 830\,G on September 3. This increase is probably due to a decay
of the penumbra beginning at the outer parts with the consequence that the remaining mean magnetic field is more
vertical.

The temporal evolution of the  total magnetic flux of umbra and penumbra is displayed in Fig.~\ref{FIG:flux}.
For both, umbra and penumbra, linear fits represent the data quite well.
The penumbra starts with a value of $100\times 10^{23}$\,Mx at the beginning of the observations and the flux 
decreases with time to $10\times 10^{23}$\,Mx on September 03. 
The umbra has a value of $10\times 10^{23}$\,Mx at the beginning and $1\times 10^{23}$\,Mx at the end of the observations. 

The temporal evolution of the magnetic flied inclination is represented in Fig.~\ref{FIG:inclination}.
For the umbra the magnetic field inclination is not changing significantly during the sunspots decay phase.
Contrarily, the inclination values for the penumbra linearly increase from around 120$^{\circ}$ to
130$^{\circ}$. This means that the magnetic field lines become more vertical in the penumbra during 
the decay of the sunspot.

%===============================================================================
%    $B_z$ at the umbral boundary
%===============================================================================

\subsection{$B_z$ at the umbral boundary}
\label{SEC4_4.2}

\textit{Jur\v{c}\'{a}k's criterion} implies that at stable umbral boundaries, the intensity contours of $0.5 I_0$ and of
$B_z = 1867$~G (in case of Hinode data) should coincide. In Fig.~\ref{FIG:jurcak1}, we show these intensity and $B_z$
contours for the eight analyzed SP scans of the decaying sunspot. Obviously, the $B_z$ contours do not spatially
coincide with the intensity contour and the \textit{Jur\v{c}\'{a}k criterion} is not valid for the analyzed decaying
sunspot.

The displacement of the green and yellow contours in Fig.~\ref{FIG:jurcak1} is increasing in time, i.e., the umbral
area with $B_z < 1867$~G is increasing during the decay phase. These umbral regions are thus prone to be transformed
into penumbra, light bridge, or granulation and the umbra is consecutively getting smaller as shown in
Fig.~\ref{FIG:area(t)}. Interestingly, during the first two/three days of the observations, the $B_z$ boundaries are not
too far off from the $0.5 I_0$ umbral boundary and the contours are matching well in certain segments of the umbral
boundaries. The intensity images show a rather stable sunspot during these first days, the decay phase just started,
and this is a possible reason for the observed partial match of the $B_z$ and intensity boundaries.

We also check if any different $B_z$ value can be found to match the $0.5 I_0$ umbral boundary during the decay
phase. In Fig.~\ref{FIG:jurcak2}, we show the best matches of intensity and $B_z$ contours resulting from this
analysis. In the first four days, we were able to identify $B_z$ values that roughly coincide with the intensity
boundaries of the umbra. The vertical component of the magnetic field at the umbral boundary of the decaying sunspot is
progressively decreasing during these four days. In the later phases of the decay, there is no unique $B_z$
value corresponding to the intensity boundary of the umbra. Interestingly, the $B_z$ value matching at least partially
the intensity boundary is higher than during the first days of the sunspot decay. 
In this study, we cannot clarify if the found
$B_z$ values are of any significance or if they are a unique property of the studied sunspot.
%###############################################################################
%#
%#    Discussion and Conclusions
%#
%##############################################################################

\section{Discussion and conclusions}
\label{SEC5}

In contrast to previous statistical investigations \citep{bumba1963, vmp_etal1993, petrovay1997}, 
we observed only a single sunspot, but under constant conditions for all observing days. For the mean, 
sign-independent vertical component of the magnetic field, we find a linear decrease in the umbra and 
an almost linear increase in the penumbra. We explain this increase by a less inclined mean magnetic 
field of the penumbra (c.f., Fig.~\ref{FIG:inclination}) when the decay sets in at the outer boundary 
and the outer, more horizontal 
magnetic field disappears first. \citet{BellotRubio2008} and \citet{Watanabe2014} suggested 
that the penumbral fields become more vertical during a sunspot decay. For the area of the penumbra 
we find an accelerated decay where the quadratic coefficient is negative indicating a negative 
curvature. On the contrary, a positive curvature was found by \cite{petrovay1997}. Towards the end, the decay does 
not show a significant deviation from a linear behavior. Although the umbral area is already reduced from the first to the second day of our observations, 
the total area is still growing. At the end of our observations, obviously new magnetic flux emerged and 
enlarged the umbral area again (see the last panel of Fig.~\ref{FIG:jurcak1}). 
By solving a diffusion equation with constant magnetic eddy diffusivity 
\citep{Meyeretal1974}, \citeauthor{Stix2002} (\citeyear{Stix2002}, p.352) explained a linear decay
of the magnetic flux, but in such models the decay curve of the umbral area is always convex as its 
second derivative is negative, $\rm{d}^2 A/\rm{d} t^2<0 $ \citep{KandR1975}.
A linear decay of area and magnetic flux can be obtained by inclusion of a weak magnetic 
quenching of the diffusivity by the (decaying) magnetic field \citep{RandK2000}. This option is 
in agreement with our results for the umbra, however, for the penumbra we see a convex decay of the area.

During the decay phase of the studied sunspot, the \textit{Jur\v{c}\'{a}k criterion} is not fulfilled 
at the UP boundary. Thus, the criterion is not valid for the decaying sunspot under study. This 
finding can be used to check if a spot already entered its decay phase even if it still appears 
morphological regular. Furthermore, we could not find any other constant value of the vertical 
magnetic field that would represent the UP boundary in decaying sunspot. 

We conclude that during the decaying phase of the sunspot, its magnetic field is getting weaker and consequently more vertical. 
The decrease of its vertical magnetic field is in the studied case faster than the time scale for the mode transition of the magnetoconvection.
Thus, in this decaying sunspot, umbral regions with $B_z < 1867$~G exist and are temporarily not occupied by penumbra or granulation. After some time, the penumbra or granulation protrude deeper into these umbral regions Consequently, the umbra gets smaller with time during the decay phase.

\begin{acknowledgements}
We express our thanks to Prof. Dr. G\"unther R\"udiger for helpful discussions and his comments to improve the manuscript.
This work was supported by the project VEGA 2/0004/16. 
\textit{Hinode} is a Japanese mission developed and launched by ISAS/JAXA, with NAOJ as domestic partner and NASA and STFC (UK) as international partners. It is operated by these agencies in cooperation with ESA and NSC (Norway). This article was created by the realisation of the project ITMS No. 26220120029, based on the supporting operational Research and development
program financed from the European Regional Development Fund.
\end{acknowledgements}

%###############################################################################
%#
%#    BIBLIOGRAPHY
%#
%###############################################################################

%\input{bibliography.tex}
\bibliographystyle{aa}
\bibliography{aa-jour,benko}

\end{document}